\documentclass[aps,prl,twocolumn,showpacs,groupeaddress]{revtex4}  

\usepackage{graphicx}  
\usepackage{bm}        
\usepackage{amssymb}   

\newcommand{\bx}{{\mathbf{x}}}

\newcommand{\bz}{{\mathbf{z}}}
\newcommand{\bxi}{{\mathbf{x_i}}}

\newcommand{\beq}{\begin{eqnarray}}
\newcommand{\eeq}{\end{eqnarray}}
\newcommand{\be}{\begin{equation}}
\newcommand{\ee}{\end{equation}}

\begin{document}

\title{Significance analysis and statistical mechanics: an application
to clustering}

\author{Marta {\L}uksza$^1$, Michael L\"assig$^2$, and Johannes Berg$^2$}
\affiliation{$^1$Max Planck Institute for Molecular Genetics, Ihnestra{\ss}e 63-73, 14195 Berlin, Germany\\
$^2$Institut f\"ur Theoretische Physik, Universit\"at zu K\"oln, Z\"ulpicher Stra{\ss}e 77, 50937 K\"oln, Germany\\}
\date{\today}

\begin{abstract}
  This paper addresses the statistical significance of structures in
  random data: Given a set of vectors and a measure of mutual
  similarity, how likely does a subset of these vectors form a cluster
  with enhanced similarity among its elements? The computation of this
  \textit{cluster $p$-value} for randomly distributed vectors is
  mapped onto a well-defined problem of statistical mechanics. We
  solve this problem analytically, establishing a connection between
  the physics of quenched disorder and multiple testing statistics in
  clustering and related problems. In an application to gene
  expression data, we find a remarkable link between the statistical
  significance of a cluster and the functional relationships between
  its genes.
  \end{abstract}

\maketitle

Clustering is a heavily used method to group the elements of a large
dataset by mutual similarity. It is usually applied without
information on the mechanism producing similar data vectors. Any
clustering depends on two ingredients: a notion of similarity between
elements of the dataset, which leads to a \textit{scoring function}
for clusters, and an \textit{algorithmic procedure} to group elements
into clusters. Diverse methods address both aspects of clustering:
similarities can be defined by Euclidean or by information-theoretic
measures~\cite{Slonim2005}, and there are many different clustering
algorithms ranging from classical $k$-means~\cite{kmeans} and
hierarchical clustering~\cite{ward1963} to recent message-passing
techniques~\cite{Frey2007}.

An important aspect of clustering is its \textit{statistical
  significance}, which poses a conceptual problem beyond scoring and
algorithmics.  First, we have to distinguish ``true'' clusters from
spurious clusters, which occur also in random data.  An example is the
starry sky: true clusters are galaxies with their stars bound to each
other by gravity, but there are also spurious constellations of stars
which are in fact unrelated and may be far from one another. Second,
clustering procedures generally produce different and competing
results, since their scoring function depends on free parameters. The
most important scoring parameter weighs number versus size of clusters
and is contained explicitly (e.g., the number $k$ in $k$-means
clustering) or implicitly (e.g., the temperature in
superparamagnetic~\cite{BlattWisemanDomany:1996} and information-based
clustering~\cite{Slonim2005}) in all clustering procedures. Choosing
smaller values of $k$ will give fewer, but larger clusters with lower
average similarity between elements. Larger values of $k$ will result
in more, but smaller clusters with higher average similarity. None of
these choices is a priori better than any other: both tight and loose
clusters may reflect important structural similarities within a
dataset.
\begin{figure}[tb]
\label{fig:data_models}
\includegraphics*[]{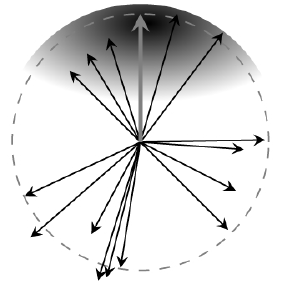}
\caption{{\bf Clustering a set of random vectors.} In a set of
  randomly chosen vectors, subsets of vectors can arise whose elements
  share a large similarity among each other. Here a cluster is shown
  with its center of mass pointing upwards and the shading indicating
  score contributions. Large clusters with high similarity among its
  elements occur only in exponentially rare configurations of the random
  vectors.  }
\end{figure}

Addressing the cluster significance problem requires a statistical
theory of clustering, which is the topic of this paper. Our aim is not
to propose a new method for clustering, but to tell significant
clusters from insignificant ones. The key result of the paper is the
analytic computation of the so-called \textit{cluster $p$-value}
$p(S)$, defined as the probability that a random data set contains a
cluster with similarity score larger than $S$. This result provides a
conceptual and practical improvement over current methods of
estimating $p$-values by simulation of an ensemble of random data
sets, which are computationally intensive and, hence, often omitted in practice.  

Our approach is based on an intimate connection between cluster statistics
and the physics of disordered systems. The score $S$ of the
highest-scoring cluster in a set of random vectors is itself a random
variable, whose cumulative probability distribution defines the
$p$-value $p(S)$. For significance analysis, we are specifically
interested in the large-$S$ tail of this distribution. Our calculation
employs the statistical mechanics of a system whose Hamiltonian is
given by (minus) the similarity score function. In this system, $\log
p(S)$ is the entropy of all data vector configurations with energy
below $-S$. We evaluate this entropy in the thermodynamic limit where
both the number of random vectors, and the dimension of the vector
space are large. In this limit, the overlap of a data vector with a
cluster center is a sum of many variables; the resulting thermodynamic
potentials can then be expressed in terms of averages over Gaussian
ensembles. 

High-scoring clusters have to be found in each
\textit{fixed configuration} of the random data vectors, which act as
quenched disorder for the statistics of clusterings. The disorder
turns out to generate correlations between the scores of clusters
centered on different directions of the data vector space. These
correlations, which become particularly significant in
high-dimensional datasets, show that clustering is an intricate
multiple-testing problem: spurious clusters may appear in many
different directions of the data vectors.  Here, we illustrate our results by
application to clustering of gene expression data, where
high-dimensional data vectors are generated by multiple measurements
of a gene under different experimental conditions. The link between quenched
disorder and multiple testing statistics is more generic, as discussed
in the conclusion.

\paragraph{Distribution of data vectors and scoring.}
We consider an ensemble of $N$ vectors $\bx_1, \bx_2, \ldots, \bx_N$, which are drawn independently from a distribution $P_0(\bx)$. We are specifically interested in data vectors with a large number of components, $M$. Clusters of such vectors are generically supported by multiple vector components, which is the source of the intricate cluster statistics discussed in this paper. We assume that the distribution $P_0 (x)$ factorizes in the vector components, 
$P_0(\bx) = p_0(x_1) \dots p_0 (x_M)$  
(this assumption can be relaxed, see below). Such null models are, of course,
always simplifications, but they are useful for significance estimates in empirical data (an example is $p$-values of sequence alignments~\cite{KarlinAltschul}).

A subset of these  vectors forms a cluster. The clustered vectors are distinguished by their mutual similarity, or equivalently, their similarity to the center $\bz$ of
the cluster, see Fig.~1.  We consider a simple similarity measure of vectors,
the Euclidean scalar product: each vector $\bx$ contributes a score
\be
\label{eq:score}
s (\bx | \bz, \mu) = \frac{1}{\sqrt{M}} \bx \cdot \bz - \mu \ .   
\ee
The scoring parameter $\mu$ acts as a threshold; vectors $\bx$ with an insufficient overlap with the cluster center $\bz$ result in a negative score contribution. The squared length of cluster centers is normalized to $\bz \cdot \bz = M$. 

A cluster can now be defined as a subset of positively scoring vectors. The {\it cluster score} is the sum of contributions from vectors in the cluster, 
\be
S(\bx_1,\ldots,\bx_N | \bz, \mu) = \sum_{i=1}^N \max\left[s(\bxi | \bz, \mu),0 \right] \ .
\label{eq:cluster_score}
\ee
Large values of $\mu$ result in clusters whose elements have a large
overlap, small values result in more loose clusters. The total score
is determined both by the number of elements and by their similarities
with the cluster center, that is, tighter clusters with fewer elements
can have scores comparable to those of looser but larger clusters. Both the
direction $\bz$ and width parameter $\mu$ of clusters are a priori unknown.

\paragraph{Cluster score statistics.} To describe the statistics of an arbitrary cluster score
$S(\bx_1,\ldots,\bx_N)$ for vectors drawn independently from the
distribution $P_0 (\bx)$, we consider the partition function 
\begin{eqnarray}
Z(\beta) & = & 
  \prod_{i=1}^{N} \int d \bx_i \, P_0(\bx_i)    
\, e^{\beta S(\bx_1, \dots, \bx_N)} 
\nonumber \\
& = & \int dS \, p(S)  \, e^{\beta S} \ . 
\label{generating_function}
\end{eqnarray}
The second step collects all configurations of vectors $(\bx_1, \dots, \bx_N)$ with 
cluster score $S$,  so $p(S)$ denotes the density of states as a function of score $S$. 
Asymptotically for large $N$, this density can be extracted from $Z(\beta)$ as  
\be
\label{eq:p}
\log p(S) \simeq N \Omega (s) - \frac{1}{2} \log (g N).
\ee
Here $\Omega (s)$ is the {\em entropy} as a function of the score  per element, $s \equiv S/N$, which is the Legendre transform of the reduced free energy density $\beta f(\beta) = - \log Z(\beta) /N$, i.e., $\Omega (s) = -\max_\beta [\beta f(\beta) + \beta s] \equiv -\beta^* f(\beta^*) - \beta^* s$. The prefactor $g$ of the subleading term is given by $g =2 \pi | (\partial^2/\partial \beta^2) \beta f(\beta)|_{\beta = \beta^*}$. The $p$-value of a cluster score $S$ is defined as the probability $\int_{S}^{\infty} dS' \, p(S')$ to find a score larger or equal to $S$. Inserting~(\ref{eq:p}) shows that this $p$-value equals $p(S)$ up to a proportionality factor of order one.

\paragraph{Clusters in a fixed direction.}
As a first step, and to illustrate the generating function~(\ref{generating_function}),  we
compute the distribution of scores for clusters with a fixed center
$\bz$. We assume that the null distribution $p_0$ for vector components has finite moments, set the first two moments to 0 and 1 without loss of generality, and we choose $\bz$ to lie in some direction which has non-zero overlap with a finite fraction of all $M$ directions. Hence,  the overlap
$x_i \equiv \bx_i \cdot \bz$ is approximately Gaussian-distributed by
the central limit theorem. The generating
function~(\ref{generating_function}) gives
\be
\label{eq:free_energy_fixed}
-\beta f_c(\beta,\mu) =
\log \left[\left(1-H\left(\mu\right)\right)+ e^{\frac{\beta^2}{2}-\beta \mu} H\left(\mu - \beta\right) \right],
\ee
where the index $c$ denotes evaluation for a fixed cluster center and
$H(x)=\int_{x}^\infty dx\, G(x)$ is the
cumulative distribution function of the Gaussian $G(x)=\exp (-
x^2/2)/\sqrt{2\pi}$. The result is
an integral over the component $x \equiv \bx \cdot \bz$ of a data
vector in the direction of the cluster center: Below the score
threshold $\mu$, the component gives zero score, which contributes the
cumulative distribution $\int_{-\infty}^{\mu} \! dx \,G(x)$ to the
partition function.  Above the score threshold, the component gives a
positive score, which generates a contribution of $\int^{\infty}_{\mu}
\!  dx \, G(x)\exp\{ \beta s(x|\mu) \}$. 
The resulting score distribution is given by~(\ref{eq:p}), $\log
p_c(S)= N \Omega\left(s=S/N\right)-(1/2) \log (g_c N)$, see Fig.~2(a).

\begin{figure}[t]
\label{fig:solution}
\includegraphics*[]{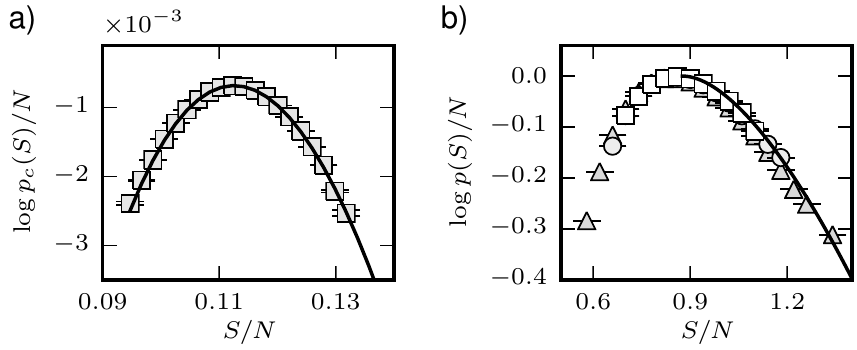}
\caption{{\bf Cluster score distributions in random data for fixed and optimal cluster direction.} Analytical distributions $p(S)$ (solid lines) are plotted against the score per element, $s = S/N$, and are compared to normalized histograms obtained from numerical experiments with $10^{6}$ samples (symbols). 
(a) Distribution $p_c (S)$ of the cluster score (\ref{eq:cluster_score}) for fixed cluster center and datasets of  $N=6000$ vectors with $M=70$, with parameter $\mu =0.1 \sqrt{M}$. Error bars show the standard error due to the finite size of the sample. 
(b) Distribution of the maximum cluster score
(\ref{eq:max_clust_score}) with parameter $\mu = 0.1 \sqrt{M}$ for
$N=40$ (triangles), $N=80$ (circles) and $N=120$ (squares),  keeping
$M/N=0.5$ fixed. 
}
\end{figure}

\paragraph{Maximal scoring clusters.} 
To gauge the statistical significance of high-scoring clusters in actual datasets we need to know the distribution of the \textit{maximum cluster score} in random data. The maximum cluster score is in turn implicitly related to the \textit{optimal cluster direction} in a dataset:
for a given subset of vectors $\bx_1, \dots, \bx_k $, the maximal cluster score is reached if the center $\bz$ coincides with
the ``center of mass'', $\bx_{\rm av} = (\bx_1 + \dots +
\bx_k)/k$.
However, 
adding or removing vectors shifts the center of mass $\bx_{\rm av}$ of the cluster and changes the score of each vector. Thus, finding the maximum score for a given dataset 
\be 
\label{eq:max_clust_score}
S_{\max}(\bx_1, \dots, \bx_N | \mu)  = 
\max_\bz S(\bx_1, \dots, \bx_N | \bz, \mu)  
\ee 
is a hard algorithmic problem, in particular for large
dimensions $M$. 
We calculate the distribution of $S_{\max}$ for independent random vectors 
from the generating function~(\ref{generating_function}) with the integral representation
\be
\label{integral_representation_maxscore}
 e^{\beta S_{\max}(\bx_1, \dots, \bx_N | \mu)} =
\lim_{\beta' \rightarrow \infty} 
\left[\int d \bz \,
e^{\beta' S(\bx_1, \dots, \bx_N | \bz,\mu)} \right]^{\beta/\beta'}
\ee 
for the statistical weight of a configuration $\bx_1, \dots, \bx_N$.
For large values of the auxiliary variable $\beta'$, only directions
$\bz$ with a high cluster score $S(\bx_1, \dots, \bx_N | \bz, \mu)$
contribute to this integral over cluster directions  $\bz$, and the maximum over the cluster
score~(\ref{eq:max_clust_score}) is reproduced in the limit 
$\beta' \to \infty$. We obtain
\be
\label{eq:free_energy_max}
-\beta f(\beta,\mu) = \min_a \left [ -\beta f_c
  \left(\beta,\mu-\frac{a}{2} \right) + \frac{M}{2N} \log
  \left(\frac{a+\beta}{a} \right) \right ] \ .  
\ee 
This expression is
to be understood in the asymptotic limit $N \to \infty$ with $M/N$
kept fixed. The result~(\ref{eq:free_energy_max}) involves a variation
over $a$, which, compared to the corresponding expression
(\ref{eq:free_energy_fixed}) for fixed cluster center, generates an
effective shift $a/2$ in the score cutoff $\mu$ and an additional
entropy-like term. The calculation uses the so-called
replica-trick~\cite{MezardParisiVirasoro,Engel2001,GardnerDerrida1988},
representing the power $n=\beta/\beta'$ of the integral
in~(\ref{integral_representation_maxscore}) by a product of $n$ copies
(replicas). The calculation proceeds for integer values of $n$, and
the limit $n\to 0$ ($\beta' \to \infty$) is taken by analytic
continuation. A key ingredient is the average {\em overlap} $q =
\langle \bz \cdot \bz' \rangle /M$ between directions of different
cluster centers for the same configuration of data vectors at finite
temperature $1/\beta'$.  We find a unique ground state (i.e., $q \to
1$ for $\beta' \to \infty$) and a low-temperature expansion
\be
q = 1 - \frac{a}{\beta'} + O\left ( \frac{1}{{\beta'}^2} \right ),  
\ee 
of the average overlap, similar to the case of directed polymers in a
random potential~\cite{HuseHenley85}, which arises in the statistics
of sequence alignment~\cite{Hwa96}. Thus, the effect of center
optimization on the free energy density (\ref{eq:free_energy_max}) and
on cluster $p$-values is related to the fluctuations between
subleading cluster centers for the same random dataset.

This solution determines the asymptotic form of the distribution of
maximum cluster score $S_{\max} = S$ as given by~(\ref{eq:p}), $ \log
p(S) = N\Omega (s) + O(\log N)$. Fig.~2(b) shows this result together
with numerical simulations for several values of $M$ and $N$, producing
good agreement already for moderate $N$. According to
(\ref{eq:free_energy_max}), the effect of center optimization on score
statistics increases with 
$M$ and
decreases with 
$N$. For small $M/N$, we
expand the solution in $N$ for fixed large $M$ and obtain $-\beta
f(\beta, \mu) = -\beta f_c (\beta, \mu) + (M/2N) \log N + {\rm
  const.}$, which leads to a distribution of maximum cluster scores
\be \log p(S) = \log p_c (S) + \frac{M}{2} \log N = N
\Omega_c(s) + \frac{M - 2}{2} \log N 
\ee 
up to terms of order $N^0$. We have generalized this calculation to
null distributions $P_0$ with arbitrary correlations between vector
components $x^1,\ldots,x^M$
~\cite{inpreparation}.

The free energy density~(\ref{eq:free_energy_max}) was derived under
the assumption of replica-symmetry (RS)\cite{MezardParisiVirasoro},
implying that only a single direction $\bz$ yields the maximal
score. This is appropriate for high-scoring clusters, since they occur
in \textit{exponentially rare} configurations of the random vectors,
for which a second cluster direction with the same score would be even
more unlikely. On the other hand, RS is known to be violated in the
case $\beta=0$, which describes clusters in \textit{typical}
configurations of the random vectors. This case has been studied
before in the context of unsupervised learning in neural
networks~\cite{Engel2001}. RS is also likely to be broken for
$\beta<0$, which describes configurations with score maxima biased
towards values lower than in typical configurations. The limit $\beta
\to -\infty$ is relevant to the problem of sphere packing in high
dimensions, for which currently only loose bounds are known.

\paragraph{Application to clusters in gene expression data.}
Clusters with high statistical significance may contain elements with a common mechanism causing their similarity. 
\begin{figure}[t]
\includegraphics*{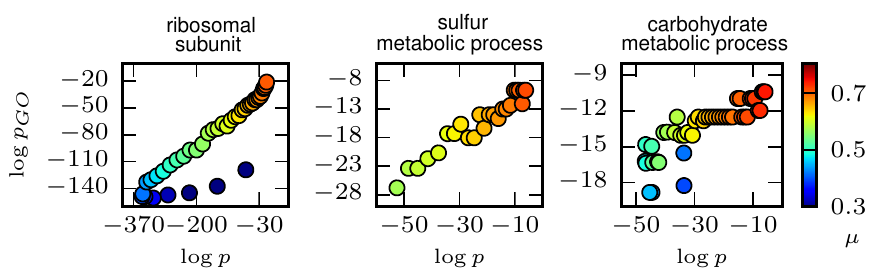}
\label{fig:correlation}
\vspace*{-.7cm}
\caption{{\bf Statistical significance of clusters correlates with functional annotation for yeast expression data}. The significance 
$p_{\rm GO}$ of gene annotation terms vs. the cluster score
significance, traced over a range of scoring parameter $\mu$ (shown by
color-scale) of three representative clusters involved in translation
(ribosomal genes), sulfur metabolic process and carbohydrate metabolic
process. 
}
\end{figure}
Here we test the link between our  $p$-value and biological function of clusters
in a dataset  of gene expression in yeast~\cite{Gasch2000, footnote}. We trace several high-scoring clusters over the range of
$\mu$ where they give a positive score. As $\mu$ increases, the
cluster opening-angle decreases (see Fig.~1), leading to a tighter,
smaller cluster.  The cluster $p$-value also
changes continuously, and the genes contained in the cluster also
change. We ask if specific functional annotations (gene ontology
GO-terms) appear repeatedly in the genes of a cluster, and how likely
it is for such a functional enrichment to arise by chance.  We compute
the $p$-value $p_{\rm GO} (C)$ of the most significantly enriched
GO-term in a cluster $C$, using parent-child enrichment
analysis~\cite{Grossman2007} with a Bonferroni correction. A cluster
with small $p_{\rm GO} (C)$ is thus significantly enriched in at least
one GO-annotation, which points to a functional relationship between
its genes. As shown in Fig. 3, the parameter dependence of the cluster
score significance $p(S(C))$ and the significance $p_{\rm GO} (C)$ of
gene annotation terms is strikingly similar. The statistical measure
based on cluster score $p$-values thus is a good predictor of
functional coherence of its elements.

\paragraph{Conclusions.}

We have established a link between quenched disorder physics and the
multiple testing statistics in clustering. This connection
applies to a much broader class of problems, which involve the parallel testing of an exponentially
large number of hypotheses on a single dataset. Examples include
imaging data (e.g. fMRI) and the analysis of next-generation
sequencing data. 
If the scores of different hypotheses are correlated with each other, the
distribution of the maximal score is not described by a known
universality class of extreme value statistics. It may still be computable by the methods used here: 
the state space of the problem is the set of all hypotheses tested (here the centers and widths of all
clusters), and configurations of data vectors generated by a null model act as quenched random disorder.

\begin{acknowledgments} Many thanks to M. Vingron for discussions
  and B. Nadler, M. Schulz, and E. Szczurek for their comments on the manuscript. This work was supported by Deutsche
  Forschungsgemeinschaft (DFG) grants GRK1360 (M. {\L}uksza), BE
  2478/2-1 (J. Berg) and SFB 680.
\end{acknowledgments}


\begin{thebibliography}{}
\bibitem{footnote} The dataset contains expression levels
  from $173$ samples for $N=6152$ genes. Raw expression levels were
  log-transformed and mean-centered, first by gene (setting the
  average expression level of a gene to zero) and then by sample
  (setting the average expression level in a sample to zero).
 Since expression levels may be correlated across
  samples (for example, in successive expression levels of a time
  course), we perform a principal component analysis~\cite{pca}. We
  restrict our analysis to the leading $M=70$ eigenvectors of the the
  yeast dataset, which account for over $95 \%$ of the gene expression
  variance.
\end{thebibliography}

\end{document}